\documentclass[preprint]{aastex}
\usepackage{color}

\def\gsim{\mathrel{\raise0.35ex\hbox{$\scriptstyle >$}\kern-0.6em
\lower0.40ex\hbox{{$\scriptstyle \sim$}}}}
\def\lsim{\mathrel{\raise0.35ex\hbox{$\scriptstyle <$}\kern-0.6em
\lower0.40ex\hbox{{$\scriptstyle \sim$}}}}

\begin{document}

\title{Matter Distribution around Galaxies}

\author{Shogo Masaki\altaffilmark{1}, Masataka Fukugita\altaffilmark{2,3,4}, Naoki Yoshida\altaffilmark{4}}

\altaffiltext{1}{Department of Physics, Nagoya University, Chikusa, Nagoya 464-8602, Japan; masaki@a.phys.nagoya-u.ac.jp}
\altaffiltext{2}{Institute for Advanced Study, Princeton, NJ08540, U. S. A.}
\altaffiltext{3}{Institute for Cosmic Ray Research, University of Tokyo, Kashiwa 277-8582, Japan}
\altaffiltext{4}{Institute for the Physics and Mathematics of the Universe, University of Tokyo, Kashiwa 277-8583, Japan}

\begin{abstract}

  We explore the mass distribution of material associated with
  galaxies from the observation of gravitational weak lensing for the
  galaxy mass correlation function with the aid of $N$-body
  simulations of dark matter. The latter is employed to unfold various
  contributions that contribute to the integrated line of sight mass
  density.  We conclude that galaxies have no definite edges of the matter
  distribution, extending to the middle to neighbouring galaxies with
  the density profile roughly $r^{-2.4}$ beyond the virial radius.  The
  mass distributed beyond the virial radius (gravitationally bound
  radius) explains the gap seen in the mass density estimates, the
  global value $\Omega_m\sim 0.27$ and typically $\Omega_{\rm gal}
  \sim 0.15$ from the luminosity density multiplied by the mass to
  light ratio.  We suggest to use a physical method of gravitational
  lensing to characterise galaxy samples rather than characterise them
  with photometric means.

\end{abstract}

\keywords{}

\section{Introduction}

Where and how is matter distributed in the Universe is a long-standing
question.  The widely accepted view is that matter of the universe is
preponderantly borne by galaxies. This leads to an estimate of the
mass density of the Universe (Ostriker et al. 1974). With modern
parameters, the luminosity density ${\cal L}=2.2\pm0.3\times 10^8h
L_\odot {\rm Mpc}^{-3}$ (Blanton et al. 2001;2003) multiplied by the
average mass to light ratio of galaxies, $\langle M/L\rangle\approx
(170\pm50)hM_\odot/L_{r\odot}$, from the gravitational lensing shear
around $2\times 10^4$ galaxies \citep{mckay01} leads to the mass
density of the Universe $\rho_m={\cal L}\times \langle M/L \rangle$
as,
\begin{equation} 
\Omega_m=0.13\pm0.05 
\label{eq:omegamgal}
\end{equation}
in units of the critical mass density.  The mass of $M/L$ we adopted
here is supposed to represent the virial mass, i.e., the mass of
particles gravitationally bound in galaxies. This $M/L$ is consistent
among a number of estimates for individual galaxies (e.g., Bahcall \& Fan 1998).

On the other hand, we now have a convincing estimate of the global matter
density from the cosmic microwave background radiation anisotropy \citep{wmap7}, 
which converges to 
\begin{equation}
\Omega_m=0.27\pm0.03.
\label{eq:omegam}
\end{equation}
The difference between Eqs. (\ref{eq:omegamgal}) and  (\ref{eq:omegam}) 
raises the problem as to where is half the matter missing,
indicated by
the mismatch of the two numbers.

Within the bound radius of galaxies, the mass distribution is inferred
to obey the profile close to the one advocated by Navarro, Frenk \&
White (1997, hereafter NFW) in the CDM dominated universe.
Simulations of clustering of dark matter conveniently point towards
this distribution for mass assemblies. This is also supported by
gravitational lensing analysis for bright galaxies
\citep{mandelbaum06}. Little is studied, however, concerning the mass
distribution beyond the virial radius.

Observational advancement with large galaxy samples lends
gravitational weak lensing to a powerful tool to explore the average
surface mass density of galaxies. The measure is
the surface density $\Sigma(R)$ as a function of the projected radius
from the centre of the  galaxy $R$,
\begin{equation}
\Sigma(R)=\int\rho\left(r=\sqrt{R^2+\chi^2}\right)d\chi,
\label{eq:sigmam}
\end{equation}
where $\rho$ is the density at $r$ and $\chi$ is  the line of sight distance,
both measured from the centre of the galaxy.
This quantity  can be explored along the line-of-sight towards a light source
by measuring its tangential shear $\gamma_{\rm t}$ or
magnification $\mu$ of the image of background sources, as
\begin{eqnarray}
\gamma_{\rm t}(R)&=&\frac{\bar\Sigma(<R)-\Sigma(R)}{\Sigma_{\rm cr}}
=\frac{\Delta\Sigma(R)}{\Sigma_{\rm cr}},\label{shear} \\
\mu(R)&\simeq&1+2\kappa(R)=1+2\frac{\Sigma(R)}{\Sigma_{\rm cr}},\label{magn}
\end{eqnarray}
where $\bar\Sigma(<R)$ is the average of $\Sigma$ within $R$, $\kappa$
is the convergence, and $\Sigma_{\rm cr}= (c^2/4\pi G)
(D_s/ D_{\ell s}D_\ell)$ is the critical surface density.  
\footnote{Eq. (5) is the Taylor expansion to the first order
in distortion. 
    M\'enard et al. (2003) examined the accuracy of the formula with
the  conclusion
    that the higher order terms contribute by about 10-15 percent to
    magnification.}
With the modern large samples this is applicable even to the
outskirt of galaxies significantly beyond the virial radius, say to $\sim
10h^{-1}$ Mpc 
using either lensing shear (Sheldon et al. 2004; Mandelbaum et al. 2006;
Reyes et al. 2010)
or statistical magnification (M\'enard et al. 2010).
While thus obtained $\Sigma(R)$
includes the surface mass density from 
neighboring galaxies
along the line of sight, this quantity gives 
a cogent information as to the
surface mass density profile of the galaxy away from its central region.

\citet[hereinafter MSFR]{MSFR}
have shown that $\Sigma(R)$ decreases approximately as
$R^{-\beta}$ with $\beta=1\pm 0.2$ to $R<1 h^{-1}$Mpc, which somewhat
flattens to $\beta= 0.6\pm 0.4$ beyond $R\approx 1h^{-1}$Mpc to
$10h^{-1}$Mpc using galaxy samples of the Sloan Digital Sky Survey
\citep{york2000,DR3} . They used the correlation of magnification
signal of background quasars with foreground galaxies. They also
showed that the surface density profile derived from the magnification
agrees with that from the tangential shear of galaxies
\citep{sheldon04}.  Mandelbaum et al. (2006) have also
derived the surface mass density profile for the luminous red galaxy (LRG)
sample, which lies parallel to that of MSFR for $R<1h^{-1}$Mpc, however,
with their amplitudes about 2.5 times larger than that of MFSR.
The difference of the surface mass density 
of the two profiles diminishes as the distance and the two
agree at $R> 1h^{-1}$Mpc, 
albeit with substantial errors in the measurements.

In this paper we study the distribution of matter around galaxies with
the aid of an extensive use of $N$-body simulations for the CDM
dominated universe without baryons, which have been convergent among
different simulations to a sufficient accuracy. We employ $N$-body
simulations to unfold contributions from multiple galaxies and
to interpret the finding from gravitational lensing, but also to
extrapolate the physics to the region beyond the reach of the
observation after verifying that the $N$-body simulation indeed
describes the observed surface mass density profile with an accuracy
sufficient to us.  We note that Hayashi \& White (2008) also used 
large numerical simulations to
study the dependence of the galaxy/halo-matter cross-correlation amplitude
on the galaxy mass and the halo mass. We expect that 
the state-of-the-art $N$-body
simulation for dark matter gives information that is
sufficiently reliable for the quantity where the observations would not
directly give the information, if the simulation is appropriately
constrained by the observation.  This gives insight as to the distribution
of the matter beyond the observation, and would tell us concerning
the more global distribution of matter in the Universe considerably
away from galaxies.

In section 2 we describe the $N$-body simulation.  In section 3 we
present the surface mass density profile in reference to the
observation. We discuss the mass distribution beyond the virial radius 
and notify that a caution is necessary as to the aperture when the
mass of galaxies is referred to in 
section 4. Section 5 is a summary of our analysis.

\section{Cosmological $N$-body simulations}

We use the parallelised $N$-body simulation code 
{\it Gadget-2}~\citep{gadget,gadget2} in its Tree-PM mode. Baryons are not
included.  We assume the flat universe with the cosmological parameters
$\Omega_m=0.258$ and $h=0.719$, the Hubble constant in units of
$100{\rm ~km~s^{-1}~Mpc^{-1}}$; for other parameters $n_s=0.963$ and
$\sigma_8=0.796$, following the parameters derived in the
WMAP 5-year result \citep{wmap5}.
The gravitational softening parameter is chosen to be
$\epsilon=10h^{-1}{\rm kpc}$.  We take $1024^3$ particles in a box of
comoving $200h^{-1}{\rm Mpc}$ on a side.  The mass of a dark matter
particle is $5.34\times10^8h^{-1}M_\odot$, so that we are able to
identify halos with mass a few times $10^{10}h^{-1}M_\odot$.  We set
the initial redshift at $z_i=50$ and generate the initial condition
using the second-order Lagrangian perturbation theory\footnote{The
parallelised second-order Lagrangian perturbation theory 
code \citep{nishimichi09} is provided 
by T.  Nishimichi.}\citep{scoccimarro98}, which has the
advantage that it is more accurate to generate initial conditions so
that one can set them at an epoch later than with
the conventional Zeldovich approximation; see
\citet{crocce06,jenkins10}.  The initial matter power spectrum at
$z=z_i$ is computed using the {\it CAMB} code \citep{CAMB}.

Haloes are identified in two-steps.  We select candidate groups of
dark matter particles using the friend-of-friends algorism, where we
take the linking length to be $b=0.2$.  We then apply the spherical
overdensity algorism to the candidate groups that are identified by
friend-of-friends.  We choose groups that contain at least $100$
particles.  We follow the conventional operational definition for the
pseudovirial radius ($r_v$): when the particles enclosed within some
radius give the average mass density $200 \times\rho_{\rm crit}(z)$,
they form a halo, the mass of which is given by the sum of those
particles and is referred to as the virial mass $M_v$.  Taking the
minimum number of particles in a halo to be $100$, i.e., the minimum
mass of the halo is $5.3\times 10^{10}h^{-1}M_\odot$, we identify
229,804 haloes. Table 1 shows the number of haloes we identified in
the simulation at $z=0.36$, which is the average redshift of 
the galaxy sample used by MSFR.

We confirmed that the halo mass function agrees with modern
simulations at a good accuracy, e.g., with that by Crocce et al.
(2010), but also with the analytic formula of Sheth \& Tormen (1999)
with a known slight deviation as reported by the above reference. We
fit all groups with the NFW profile and compare the concentration
parameter $c=r_{\rm vir}/r_s$ ($r_s $ is the scale length of the NFW profile)
as a function of halo mass with other modern simulations.  We
confirmed that our $c$ agrees with that of \citet{maccio08} (see
also \citet{bullock01}) within a few percent level. The simulations
are well converged and the difference in simulations and analysis
algorisms is tolerably small.

The scale range that concerns in this paper is 20$h^{-1}$ kpc or
larger, where the contribution from baryons modifies little
the surface mass density profile.   Hence, we confine ourselves to
the CDM universe without baryons.

\begin{table}
 \caption{Haloes identified in our simulation at $z=0.36$.}\vskip2mm
 \label{numfrac}
 \begin{tabular}{@{}lrrccc}
  \hline
  mass  [$h^{-1}M_\odot$] & number & number fraction [\%] & mass fraction [\%]\\\hline
  $5\times10^{10}-5\times10^{11}$ & 189,578 & 82.5 &  5.9\\
  $5\times10^{11}-5\times10^{12}$ & 36,402 & 15.8&  8.25 \\
  $5\times10^{12}-5\times10^{13}$ & 3,637 & 1.6& 7.64   \\
  $5\times10^{13}-5\times10^{14}$ & 187 & 0.08&  3.13  \\
  \hline
 \end{tabular}
\end{table}

\section{Surface mass density profiles around haloes}

We calculate the mean surface mass density around haloes in the
following way.  We shift the entire box so that it is centred on a
halo being considered.  We consider a beam and project the mass of
simulation particles within comoving $100h^{-1}{\rm Mpc}$ widths to
the beam. This gives the surface mass density around the beam at a
specific projected distance.  We then take the average of the beams
around all haloes to obtain the projected surface mass density
profile.

\begin{figure}
\plotone{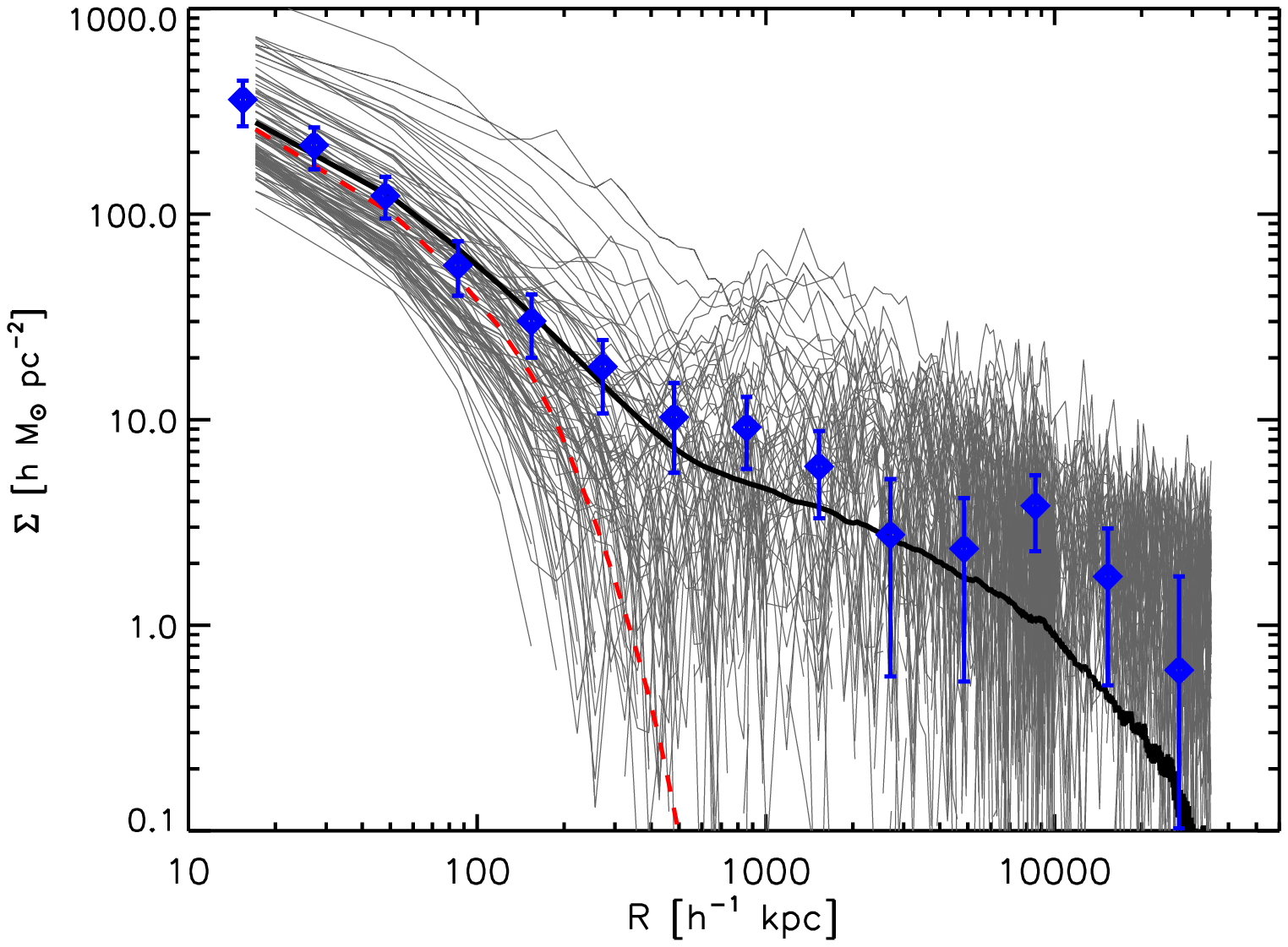}
\caption{The surface mass density profile as a function of
the distance from the centre of galaxies. The bunch of thin solid
curves represent the profiles for 100 randomly picked-up haloes from
9970 haloes with mass
$\ge 2\times10^{12} h^{-1}M_\odot$ genarated in  
the $\Lambda$CDM
$N$-body simulation. The thick solid curve is the mean of
all haloes above the mass threshold. 
The dashed curve shows the contribution from the one-halo term.
The data with error bars are the observational estimate
using galaxy mass correlation function deduced from
gravitational weak lensing for quasar brightness in MSFR.
The abscissa is the physical distance at $z=0.36$, which is
the average redshift of the weak lensing observation.
\label{sigma}}
\end{figure}

Figure \ref{sigma} compares our simulation for the surface mass
density at $z=0.36$ with the observational result of MSFR which was
derived at the same mean redshift.  The abscissa is the physical distance
at this redshift from the centre of the halo.  The simulation is
represented by a bunch of thin curves for 100  haloes randomly
chosen from the 9970 haloes that have a virial mass larger than
$2\times10^{12} h^{-1}M_\odot$ (this choice is discussed below).  The
maximum halo mass in our simulation is $5\times10^{14} h^{-1}M_\odot$.
The thick solid curve is the average over the entire sample above the
mass threshold.  The observational data (ticks with error bars) are
taken from MSFR.  The simulation agrees with the observation very
well, in both overall shape and amplitude up to $10h^{-1}$ Mpc.

The dashed curve shows the contribution from the central haloes which
are truncated at their virial radii, the so-called one-halo term.  The
figure shows that the mean surface density at $R \la 200h^{-1}{\rm
  kpc}$ is dominated by the one-halo term. We note that our column
integrates over all particles along the line of sight
and it would receive the contribution not only from the tail of
neighbouring haloes but also from possible \lq unbound' particles away from 
galaxies.

\begin{figure}
\plotone{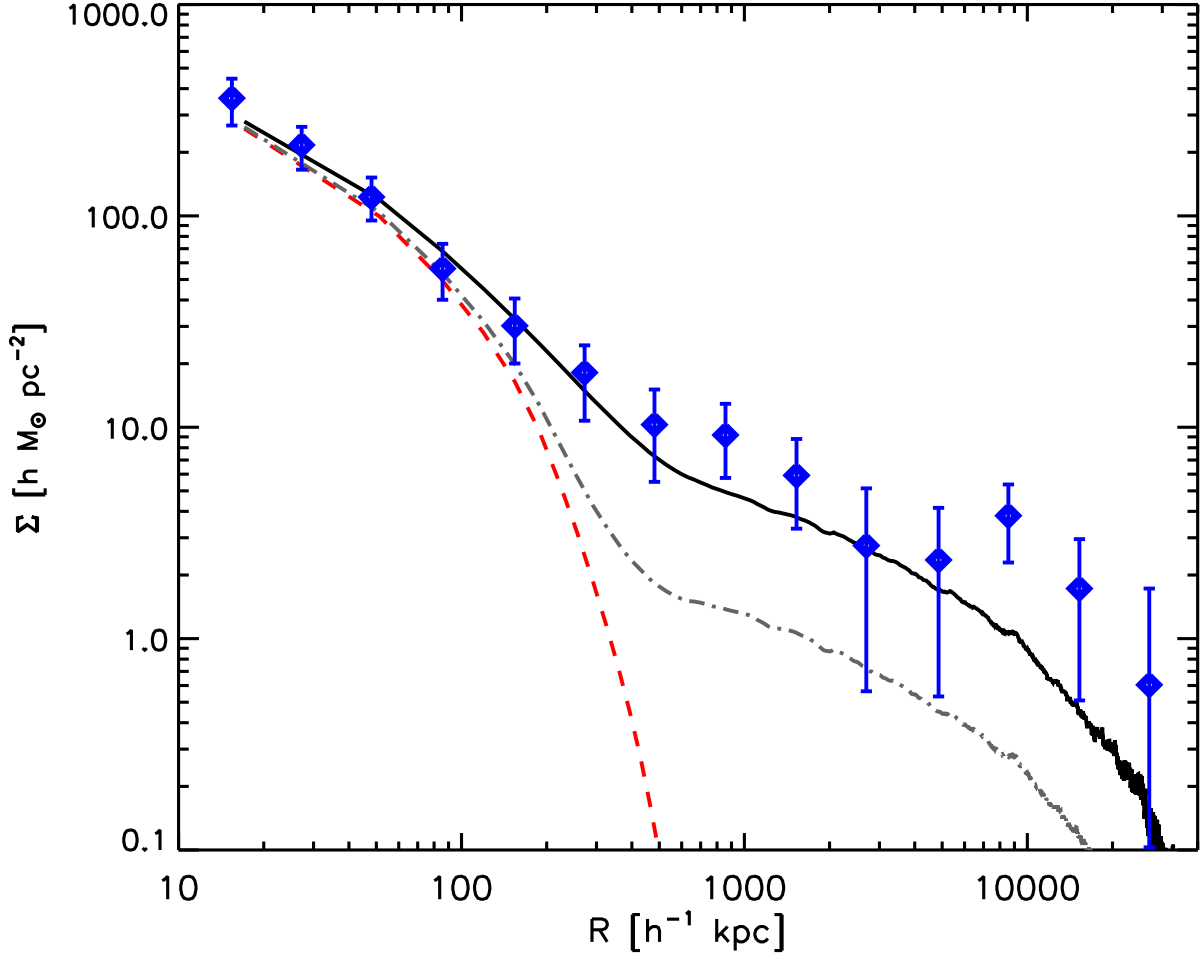}
\caption{The mean surface mass density profile as a function of
the distance from the centre of galaxies. The thick solid curve is the mean of
all haloes above the mass threshold. The dash-dotted curve represents 
the contribution from particles bound to haloes, 
i.e., particles that reside within the
virial radius of all haloes. 
The data with error bars are the observational estimate
by MSFR as in the previous figure.
\label{sigma2}}
\end{figure}

To separate the contributions of bound and unbound particles,
we remove the particles beyond the virial radii of all haloes.
Then we recompute the surface mass density in the same way 
as described above, but including haloes with mass below our virial
mass threshold set above.
This is shown in Figure \ref{sigma2} with the dash-dotted
curve, which stands for the contributions from bound particles
in all haloes. The data, the thick solid and the dashed curves are 
the same as in  Figure \ref{sigma}.
The figure shows a substantial difference between the solid and
dash-dotted curves, meaning that particles bound in haloes contribute
only by 1/3 the
total surface mass density  beyond
the distance of $\approx500h^{-1}$kpc from the galaxy. We remember that
the total surface mass density derived in the simulation 
agrees with that
estimated from gravitational lensing.  Roughly 2/3 the surface mass
density at such a distance is due to particles residing beyond the
virial radius of any galaxies, i.e., gravitationally \lq unbound' particles.

\begin{figure}
\plotone{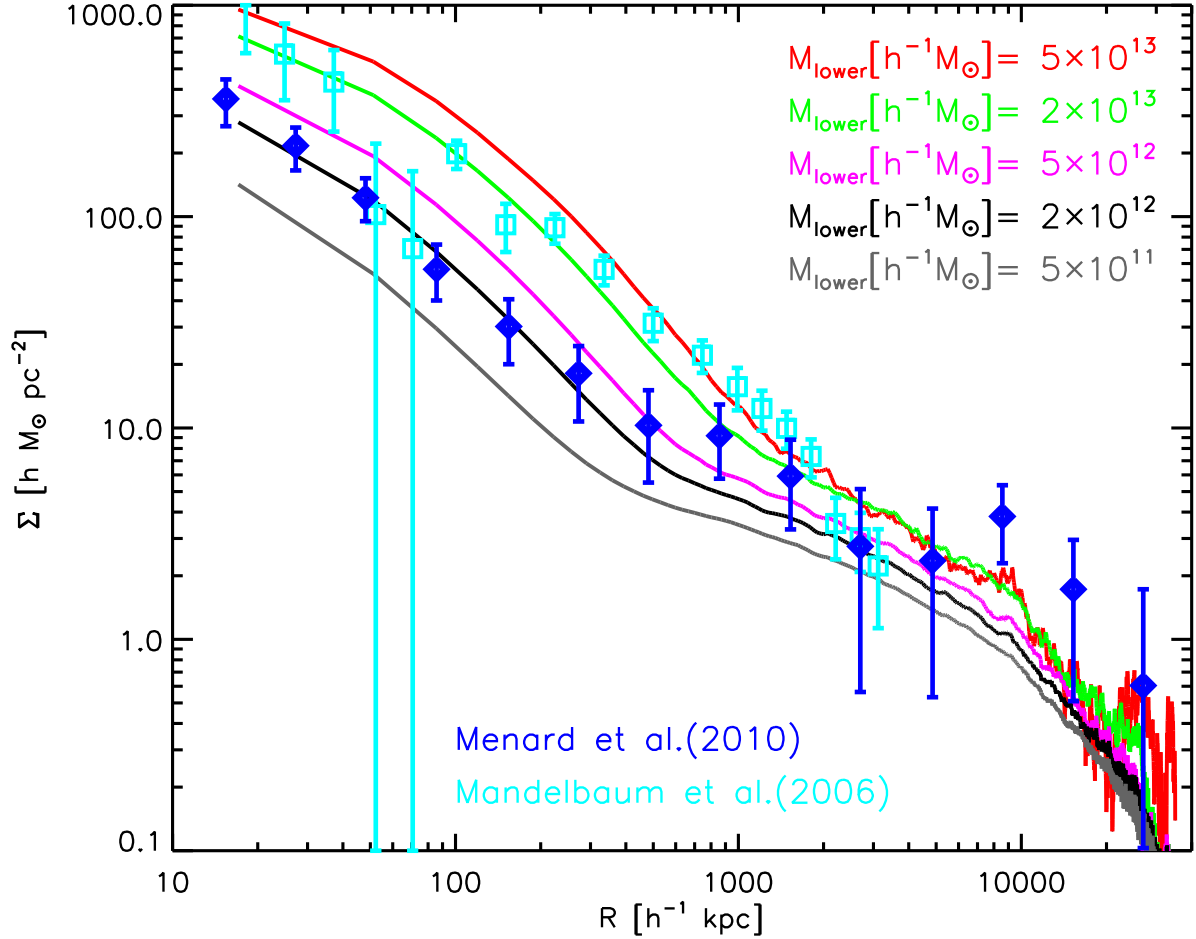}
\caption{The mean surface mass density profile as a function of
the distance from the centre of galaxies for galaxy samples with the
different threshold halo mass (virial mass). The open squares which
represent the data for LRG given by 
\citet{mandelbaum06}
are added to the MSFR data for the SDSS main galaxy sample 
shown with solid diamonds with error bars.}
\label{sigma3}
\end{figure}

In our argument we set the cutoff on the lowest mass of haloes,
somewhat arbitrarily, at $2\times10^{12}h^{-1}M_\odot$. This choice should of
course affect the surface mass density profile. We show in Figure
\ref{sigma3} the mass density profiles with different threshold mass:
$5\times10^{11},2\times10^{12},5\times10^{12},2\times10^{13},
5\times10^{13}h^{-1}M_\odot$. We see that the surface mass density
increases as the lower cutoff mass (virial mass) $M_{\rm low}$ 
increases for small radii, say,
$R<1h^{-1}{\rm Mpc}$ where the one halo term is the significant contributor.  
The result of the simulation shows $\Sigma(R)\propto M_{\rm low}^{2/3}$ for
the distance scale 10kpc to a few hundred kpc: 
it is summarised, e.g., at
120$h^{-1}$kpc as 
\begin{equation}
\Sigma\simeq100(M_{\rm low}/8\times
10^{12}h^{-1}M_\odot)^{2/3}hM_\odot({\rm pc})^{-2}\ .
\label{eq:sigma-mlow}
\end{equation} 
With the singular isothermal sphere we have the surface mass density at 
the projected distance $R$,
\begin{equation}
\Sigma(R)=\left({25\pi\over 6}\rho_{\rm crit}\right)^{1/3} M_v^{2/3}{1\over R},
\label{eq:SIS-sigma}
\end{equation}
and a similar relation holds with the NFW profile albeit in the limited
distance range.  With our mass function the average mass $\langle
M_v\rangle\approx 5M_{\rm low}$, so that our fitting formula 
eq.(\ref{eq:sigma-mlow}) gives for a given $\Sigma(R)$ a
halo mass 2.5 times smaller than the model with the singular
isothermal sphere.

This sample threshold dependence explains the difference in the surface
mass density distributions between \citet{mandelbaum06}, which are also
plotted in Figure \ref{sigma3}, and MSFR.
The former gives $\Sigma(R)$ larger by a factor of 2.5 than MSFR
within a few hundred kpc range. 
Equation (6) then indicates that the threshold mass
of the LRG sample of Mandelbaum et al. (2006) is approximately by 4 times 
more massive than that of MSFR, who used the main galaxy sample.
The difference seen in $\Sigma(R)$ among different halo masses
diminishes for a large $R$, where the one halo term no longer
dominates but $\Sigma(R)$ is contributed by neighbouring haloes and
unbound particles, as 
seen by comparing
$\Sigma(R)$ of \citet{mandelbaum06} and MSFR.
\footnote{
The LRG 
data show a slightly larger amplitude at $R \sim 1h^{-1}$ Mpc, reflecting
greater bias. The overall features of the
surface density profiles for different galaxy/halo masses 
are studied in detail by \citet{HW08}.}
(Neighbouring haloes are more likely to be 
those of normal galaxies rather than LRGs.)

Figure \ref{sigma3}
indicates that the mean surface mass density profile
around galaxies measured by MSFR is reproduced well when the threshold
mass $M_{\rm low}$ is set to $2\times 10^{12}h^{-1}M_\odot$, which is
approximately the mass of the $L^*$ galaxy $1.2\times
10^{12}h^{-1}M_\odot$ \citep{FP06}; hence our default cutoff is
chosen.  MSFR used the galaxy sample $17<i<21$. If the effective
cutoff of the sample would be around $i\approx 20$, sample's threshold
is around the $L^*$ galaxy\footnote{ At a more accurate level we must
  consider that $L^*$ luminosity corresponds to $10^{10.6}L_\odot$ in
  the $i$ band, while it does to $10^{10.5}L_\odot$ in the $r$
  band. $L^*$ luminosity is not physically well defined. We also note
  that there is a significant uncertainty in the mass-luminosity
  relation associated with morphology of galaxies.}\citep{blanton01}.

It is less ambiguous to estimate the threshold mass from $\Sigma(R)$
itself. With the threshold $2\times 10^{12}h^{-1}M_\odot$, $\Sigma(R)$
of the simulation is in good agreement with the data. If we would take
the threshold $5\times 10^{11}h^{-1}M_\odot$, which roughly
corresponds to the $i\approx 21$ mag threshold, $\Sigma(R)$ of the simulation
would lie somewhat too low.  This consideration suggests that the
surface mass density can be used to characterise the mass of the
galaxy sample, which otherwise is difficult to estimate accurately.

In the limited range of the distance scale we are studying, $15{\rm
  kpc}<R<200{\rm kpc}$ both NFW and singular isothermal sphere give
mass profiles that are very similar and both give good fits to the data,
so that  we are not able to distinguish between the two profiles in this
range.

\section{Mass distribution beyond the virial radius}

\begin{figure}
\plotone{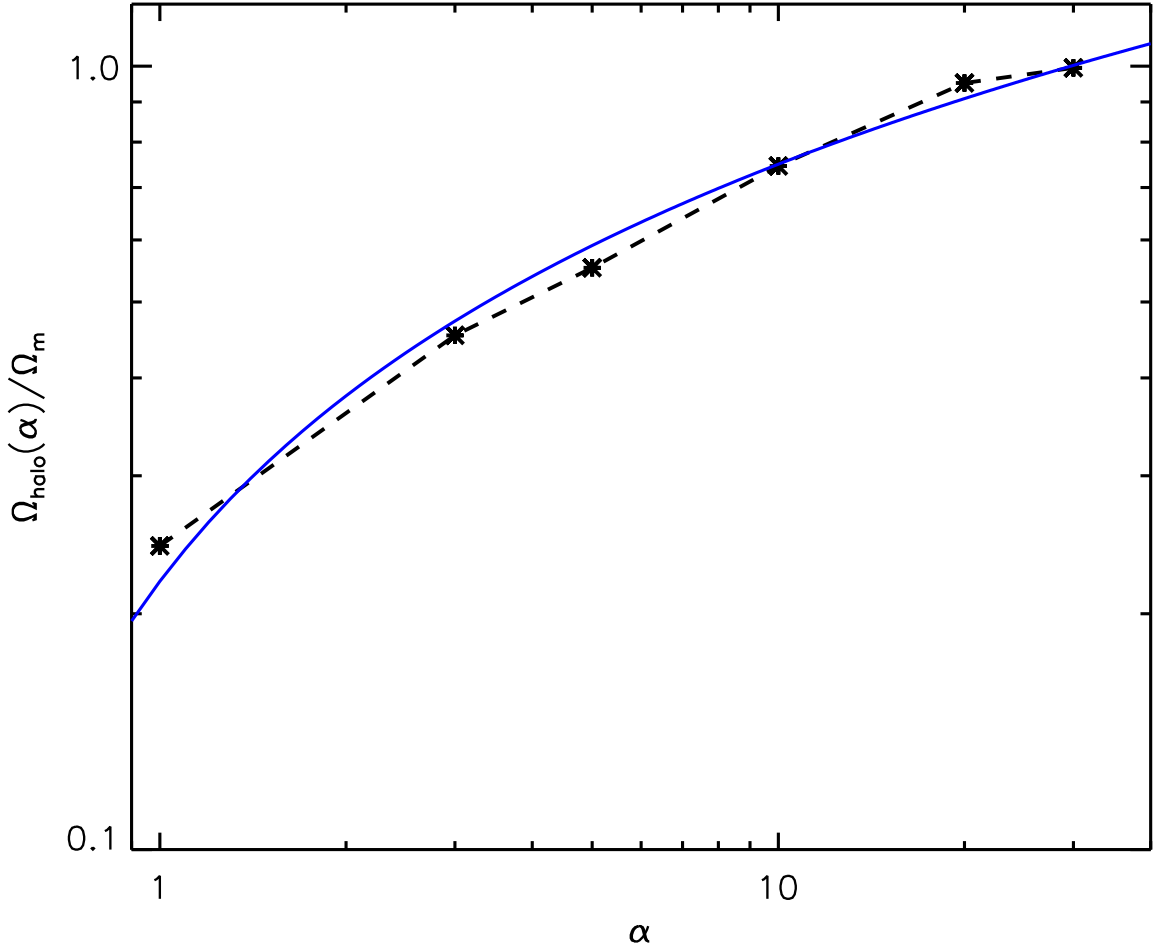}
\caption{Fraction of mass contained in the sphere centred on individual
haloes with radius $\alpha R_{\rm vir}$, where $ R_{\rm vir}$ is the pseudovirial
radius and $\alpha$ is the multiplier represented in the abscissa.
The plot is for the sample with the threshold halo mass 
$1\times 10^{11} M_{\odot}$.
The solid curve is $0.23\ln\alpha + 0.22$ given in the text.  
\label{omega}}
\end{figure}

We see above that the substantial amount of matter in the Universe
resides outside the pseudovirial radius of galaxies.  In order to
examine whether the distribution of unbound particles is organised, we
calculate the total amount of mass encircled with the pseudovirial
radius, and then expand the encircling radius by a factor of $\alpha$:
we denote it in units of the critical density as $\Omega_{\rm
  halo~extended}(\alpha)$.  The pseudovirial radius is defined for
each galaxy as 200 times $\rho_{\rm crit}$.  For a large $\alpha$ two
haloes start overlapping, and in such cases we count the amount of
material only once, avoiding double counting.  Figure \ref{omega}
shows the fraction $\Omega_{\rm halo~extended}(\alpha)/\Omega_m$ as a
function of $\alpha$, which is described well with $0.23\ln\alpha +0.22$
consistent with the NFW profile in its $\rho\sim r^{-3}$ regime.

If the mass distribution of a galaxy were spatially bounded, it would
not matter for the mass estimate of galaxies 
whatever the encircling radius is taken in so far as it
is large enough.  We find, however, this is not the case. The
encircling radius must be carefully specified, for instance, to
calculate the mass to light ratio to be quantitatively meaningful.

The figure shows that the matter within the pseudovirial radius
($\alpha=1$) is only 0.25 times the total matter if the sample of galaxies 
is set above the mass thereshold $10^{11}h^{-1}M_\odot$,
above which the galaxy mass (and hence $M/L$) is
observationally estimated; see below. This fraction (and the curve
in the figure) depends
on the  threshold mass of sample galaxies, 
as $\sim -\log M_{\rm low}$ in so far as $M_{\rm low}$ is taken below the
break mass of the mass function.
If the threshold is taken at a larger mass, 
this fraction decreases faster, mainly because the number of galaxies included 
in the sample decreases. 
Note that the sample threshold must be taken small enough to estimate
the mass density of the universe so that the
contribution of
sub-threshold galaxies to the global quantity (in practise
luminosity density) is not substantial.  
We confirm that this mass fraction agrees with the mass obtained 
by integrating over the mass
function. 
The figure also shows that this fraction
becomes close to 0.5 if 3 times the pseudovirial radius is taken to
encircle clustering.  Almost entire matter ($>90$\%) is included
only with $\alpha>20$.

Now it may be appropriate to examine the case discussed by \citet{mckay01},
who estimated the mass of galaxies from gravitational lensing shear, as
we quoted earlier.  They used the spectroscopic sample of SDSS, which
means the limiting magnitude accurately defined at $r_{\rm
  limit}=17.8$ \citep{strauss02}.  It corresponds to $M_r\approx
-20.6$ for the median redshift of the sample $z\approx 0.10$ 
with the median K correction, or to $0.41L_r^*$. If we invoke
an empirical scaling relation $M\propto L^2$ \citep{mandelbaum06} as
consistent with the Faber-Jackson relation, we
estimate the limiting mass of the sample $M_{\rm low}\approx 2\times
10^{11}h^{-1}M_\odot$, taking $M_{\rm MW}\approx 2.3\times
10^{12}M_\odot$ \citep{FP06} and $L_{\rm MW}=1.4L_*$ for 
the Milky Way to normalise the $M-L$ relation.

This can also be verified from the gravitational lensing measurement
itself.  \citet{mckay01} obtained $\Sigma(R)\approx 2.5(R/1h^{-1}{\rm
  Mpc})^{-0.8\pm0.2} hM_\odot {\rm pc}^{-2}$. Consulting with our
$\Sigma(R)$ relation with the sample threshold mass as a parameter
(see Figure \ref{sigma3}), we estimate $M_{\rm low}\approx 1.5\times
10^{11}h^{-1}M_\odot$ in agreement with the photometric estimate.

With this threshold we estimate the mean virial
mass of the spectroscopic sample approximately ${\overline
  M_v}\approx 1.5\times 10^{12}h^{-1}M_\odot$ using the 
mass function obtained in our simulation. 
This value is compared with the estimate ${\overline
  M_v}\approx2.6\pm1.3\times 10^{12}h^{-1}M_\odot$ of
\citet{mckay01} for their sample,
where the error includes the variance associated with the galaxy
morphology. This shows the consistency of the two estimates,
those by \citet{mckay01} and ours.

Our simulation gives the average pseudovirial radius over the sample 
with the lower mass cutoff $M_{\rm low}$ as
${\overline r_v}=100h^{-1}{\rm kpc} (M_ {\rm low}/10^{11}
h^{-1}M_\odot)^{0.29}$, so that our estimate of the virial radius 
$\overline{r_v}$ 
for the sample with $M_{\rm low}\approx 1.5\times 10^{11}h^{-1}M_\odot$ is
${\overline r_v}=120h^{-1}$ kpc.
\citet{mckay01} claim that they
measured the $M/L$ within $260h^{-1}$ kpc, which is $\alpha\approx
2.2$ times the mean virial radius of the sample.  
Figure \ref{omega} tells us that for $\alpha=2.2$ 
approximately 0.4 times the total $\Omega_m$ should be included in their
estimate. 
This leads us to the global matter density from galaxies to
be 
\begin{equation}
\Omega_m=(0.13\pm0.05)/0.4 \approx 0.32,
\end{equation}
which is consistent with
the global value of Eq.(\ref{eq:omegam}). Although our estimate presented
here is admittedly crude, this agreement indicates that the mass
beyond the pseudovirial radius we inferred here is probably broadly correct.  We
should underline the importance of the estimate of the radius (e.g., with
respect to the virial radius) when the mass or the mass to light ratio
is presented for galaxies, since galaxies are extended objects without
definite boundaries.

\begin{figure}
\plotone{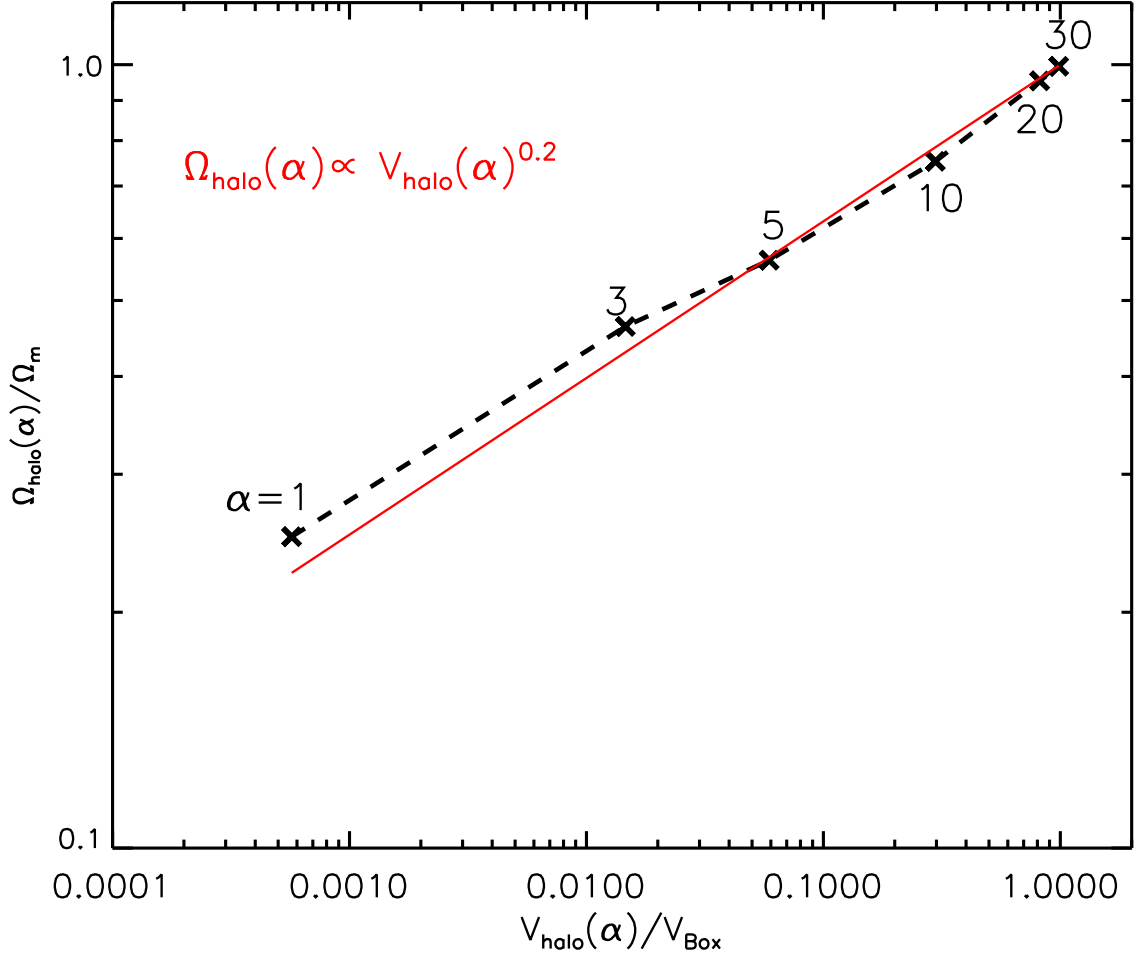}
\caption{Mass fraction borne by extended haloes with the radius 
$r\le\alpha R_{\rm vir}$, 
$\Omega_{\rm halo~extension}(\alpha)/\Omega_m$
versus the volume fraction occupied by the extension of virial spheres
$V_{\rm halo`extension}(\alpha)/V_{\rm tot}$. The numbers shown beside the
symbol is the multiplier $\alpha$.
The solid line indicates
$\Omega_{\rm halo~extension}/\Omega_m\sim (V_{\rm halo~extension}/V_{\rm tot})^{0.2}$.
\label{omegavsvol}}
\end{figure}

We next calculate the volume occupied with extended haloes as a
function of $\alpha$. In Figure \ref{omegavsvol}
we plot the fraction of mass contained in the
extension of haloes $\Omega_{\rm halo~extension}(\alpha)/\Omega_m$ as
a function of the fraction of the volume occupancy $V_{\rm
  halo~extension}(\alpha)/V_{\rm total}$, where $V_{\rm total}$ is the total
simulation volume.  The numbers beside the symbol indicates the
multiplier $\alpha$. 
$V_{\rm
  halo~extension}$ is approximately proportional to $(\alpha R)^3$ up
to overlaps of haloes at a large $\alpha$.  This figure shows that the
plot is given nearly by a straight line, indicating that
\begin{equation}
\Omega_{\rm halo~extension}(\alpha)/\Omega_m \sim \left[ V_{\rm halo~extension}(\alpha)/V_{\rm total} \right]^{0.2}
\label{eq:m-V}
\end{equation}
Let us recall if the mass distribution is random throughout the entire volume
we expect
$\Omega_{\rm halo~extension}\propto V_{\rm halo~extension}$.  The
figure shows that there is no symptom of a conspicuous break of the
curve at any $\alpha$, meaning that the distribution of unbound
matter is all organised to a distance significantly away from the
galaxy, without leaving a significant amount of material in the intergalactic
space. This power means that the matter density behaves as
\begin{equation}
\rho_m \sim r^{-2.4},
\end{equation}
beyond the virial radius, as long as halfway to the neighbouring galaxy.  This
drops faster than the isothermal profile but is numerically consistent with the
tail of the NFW profile in the range $r/r_s= 5 - 100$ with $r_s$ the NFW scale
radius. Remembering that $r_v/r_s\approx 5$ for galaxies and the
typical spacing between the galaxies is $r_0\approx 5h^{-1}$Mpc, the relevant
range matches the inter-galaxy distance.  

Since galaxies have no clear edges, it is not appropriate to define
their \lq total' mass.
If we include the mass in the tail of the galaxy halfway to the
neighbouring galaxy, the mass of galaxy increases approximately 
by a factor of 2 so
that the effective $M/L$ becomes $\approx350hM_\odot L_\odot^{-1}$. 
When multiplied by
the luminosity density ${\cal L}=2.2\times10^8hL_\odot{\rm Mpc}^{-3}$, we 
would obtain
\begin{equation}
\rho_m={\cal L}\times\langle M/L \rangle \approx 0.3\rho_{\rm crit},~~\Omega_m\approx0.3.
\end{equation}
We remark that clusters serve as a good natural integrator of the mass of
galaxies. The mass of clusters is usually estimated at the radius,
say, $r_{500}$ which is far beyond the virial radii of individual
member galaxies. Therefore mass of galaxies resided in the tail is
largely integrated in the estimate of the cluster mass. This explains
the reason why $M/L$ of clusters reaches 300$-$400, significantly
larger than the estimates for individual galaxies: the mass here
includes the mass present in the tail of galaxies.  This explains the
reason why we are arrived at the correct global mass density of the
Universe, if the {\it cluster} $M/L$ is used to multiply on the
luminosity density of {\it field} galaxies instead of $M/L$ of
individual galaxies, though this is apparently an incongruous
treatment.

\section{Summary}

We showed that the state of the art $N$-body simulation of dark matter
based on the
$\Lambda$CDM model gives an excellent description of the surface
density profile of the mass distribution around galaxies, which has
been explored up to the 10$h^{-1}$ Mpc scale from the galaxy mass
correlation function using weak gravitational lensing analysis applied
to large modern galaxy samples.  The surface profile thus derived is
consistent with $r^{-1\pm0.2}$ up to $1 h^{-1}$Mpc and somewhat
flattens to $r^{-0.6\pm0.4}$ beyond this radius. The latter power is
consistent with that of the two point correlation function of
galaxies. The galaxy mass correlation function measures the
mass distribution within the galaxy halo in a short distance scale
(smaller than a few $\times$ 100 kpc scale) and reflects the galaxy
distribution beyond this radius. The two distributions are 
similar and match each other with a slight break.

The amplitude of the surface mass density profile depends on the
galaxy sample, in particular on the lower mass cutoff applied to the
sample. Hence the amplitude serves us to infer the properties of the
galaxy sample in a self-consistent way.

We showed that the galaxy has no clear edges in the dark matter
distribution, unlike luminous matter, which should be bounded by the
cooling radius.  
The distribution is extended beyond the virial radius
in an organised way halfway to the neighbouring galaxy, so that the
Universe is filled with the material associated with tails of
galaxies, and we then call the peaks of the matter distribution
galaxies.
Inter-galactic space is filled with
  matter. Tails of galaxies extend to great distances without
  cutoff, whereas luminous component of galaxies have
  definite cutoff radius corresponding to the cooling radius.

About half the matter in the Universe is gravitationally unbound at $z\sim0$.
Its distribution, however, is never random or uniform, but is well
organised in a way to be consistent with the tail of galaxies with the
mass density roughly $\rho\sim r^{-2.4}$.  Half the matter is present
in the tail of galaxies beyond the pseudovirial radius.  This explains
the gap in the estimate of mass density of the Universe between the
global value and the value obtained by adding the contributions from
matter bound to individual galaxies: with this extended matter
distribution the matter entry closes in the mass inventory, which has
been left unclosed in \citet{FP04}\footnote{
\citet{FP04} consider that 60\% of matter is bound, i.e.,  within
virial radii, and that the rest 40\% lie outside. Their inference of these
numbers uses $M/L$ estimated by McKay et al. (2001), assuming that their
encircled radii are sufficiently large to cover the binding radius,
so that the used radius is 2.2 times the virial radius, as we discussed
in the text above. In our present estimate 60\% must be modified
to 40\% in the same context, and the bound component, if defined 
by the pseudovirial radius,
is 25\%.}.

The observables derived from gravitational lensing lend us to characterise
the sample in terms of the mass with physical means. The agreement between 
photometrically inferred characteristics and physically derived
ones is, if not perfect yet, nearly satisfactory. Gravitational lensing
would be used to characterise the sample with physical methods. 

\vskip10mm\noindent
{\bf Acknowledgement}

We thank Brice M{\' e}nard, Masamune Oguri and Rachel Mandelbaum for 
useful discussions.
We appreciate Tomoaki Ishiyama for providing us with data of their 
$N$-body simulations.
This work is supported in Nagoya and in Tokyo in part by the Grants in Aid
of the Ministry of Education.  MF acknowledges the support of the
Friends of the Institute in Princeton.

\end{document}